\documentclass[twocolumn,pra,aps,showpacs,superscriptaddress]{revtex4-1}
\usepackage[utf8]{inputenc}
\usepackage{url,psfrag,graphicx}
\usepackage{dcolumn}
\usepackage{amsmath,amssymb}
\usepackage{pstricks}
\usepackage{epsfig,placeins,float}
\usepackage{hyperref}

\def\<{\left<}
\def\>{\right>}
\def\ket|#1>{\left|#1\right>}
\def\bra<#1|{\left<#1\right|}
\def\elem<#1|#2|#3>{\left<#1\right|#2\left|#3\right>}
\def\({\left(}
\def\){\right)}

\def\pmat#1{\begin{pmatrix}#1\end{pmatrix}}
\def\{{\left\lbrace}
\def\}{\right\rbrace}
\def\beq{\begin{equation}}
\def\eeq{\end{equation}}
\def\min{\mathrm{min}}

\def\eff{\mathrm{eff}}
\def\Tr{\mathrm{Tr}}

\makeatletter
\def\inmod#1{\allowbreak\mkern5mu({\operator@font mod}\,\,#1)}
\makeatother

\begin{document}

\title[Short Title]{Entanglement detachment in fermionic systems}

\author{Hern\'an Santos}
\affiliation{Dep. de F\'{\i}sica Fundamental, Universidad Nacional de
	Educaci\'on a Distancia (UNED), Madrid, Spain} 
\affiliation{Dep. de F\'{\i}sica de la Materia Condensada, Universidad
  Aut\'{o}noma de Madrid, Cantoblanco, 28049 Madrid, Spain} 

\author{Jos\'e E. Alvarellos}
\affiliation{Dep. de F\'{\i}sica Fundamental, Universidad Nacional de
  Educaci\'on a Distancia (UNED), Madrid, Spain} 

\author{Javier Rodr\'{\i}guez-Laguna}
\affiliation{Dep. de F\'{\i}sica Fundamental, Universidad Nacional de
  Educaci\'on a Distancia (UNED), Madrid, Spain}

\date{September 18, 2018}

\begin{abstract}
  This article introduces and discusses the concept of {\em
    entanglement detachment}. Under some circumstances, enlarging a
  few couplings of a Hamiltonian can effectively detach a (possibly
  disjoint) block within the ground state. This detachment is
  characterized by a sharp decrease in the entanglement entropy
  between block and environment, and leads to an increase of the
  internal correlations between the (possibly distant) sites of the
  block. We provide some examples of this detachment in free fermionic
  systems. The first example is an edge-dimerized chain, where the
  second and penultimate hoppings are increased. In that case, the two
  extreme sites constitute a block which disentangles from the rest of
  the chain. Further examples are given by (a) a superlattice which
  can be detached from a 1D chain, and (b) a star-graph, where the
  extreme sites can be detached or not depending on the presence of an
  external magnetic field, in analogy with the Aharonov-Bohm
  effect. We characterize these detached blocks by their reduced
  matrices, specially through their entanglement spectrum and
  entanglement Hamiltonian.
\end{abstract}


\maketitle


\section{Introduction}

Quantum many-body physics constitute a theoretical crossroad between
condensed matter physics and quantum information, since both of them
benefit from the study of correlations and entanglement structures
\cite{Nielsen.00,Amico.08}. Indeed, the ground states (GS) of quantum
systems present very interesting entanglement properties. Under a
variety of circumstances, they follow the {\em area law}
\cite{Eisert.10}, which determines that the entanglement entropy of a
region is proportional to the measure of its boundary. The area law
holds for 1D gapped systems \cite{Hastings.07}, but some 1D gapless
systems are known to violate it, such as the {\em rainbow system}
\cite{Vitagliano.10,Ramirez.14,Ramirez.15}, a free fermionic chain
with exponentially decaying hoppings as we move away from the center,
which presents maximal entanglement between the left and right halves
in its GS.

In this work we will characterize systems where the increase of some
couplings leads to a block effectively {\em detaching} from the system
within its GS. Thus, as those couplings grow we observe a sharp
decline in the entanglement entropy between the block and its
environment. The block in question need not be formed by contiguous
sites. Effectively, in a first example, the block can be constituted
by the two extreme sites of an open free-fermionic chain, and the
selected couplings will be the second and penultimate hoppings
\cite{Santos.18}. A further consequence is that, through entanglement
monogamy \cite{Koashi.04}, the sites of the block become more
correlated among themselves, even if they are far away. In our two
other examples the detached block is given by a superlattice of a
chain and the edge sites of a star-graph, always for a free-fermion
system. Also in these cases, the block sites establish large
correlations, despite being far away.

When the entanglement entropy between block and environment falls to
zero the GS can be effectively factorized into two wavefunctions, one
for the block and the other for the environment. For low entanglement
entropy, but still non-zero, the best way to characterize physics on
the block is still to analyze its reduced density matrix. Indeed, we
have considered in all the relevant cases the entanglement spectrum
\cite{Li_Haldane.08} and the entanglement Hamiltonian
\cite{Peschel.03,Eisler.17,Tonni.18}. In the studied cases, we show that both
the superlattice and the edge sites of the star-graph constitute an
effective 1D free-fermionic system. The case of the star-graph
presents special relevance because whether the edge sites detach or
not depends on the number of legs of the star, modulo 4. Moreover, an
external magnetic field can re-connect the edge sites to the rest of
the system, or detach them again, in a way which is reminiscent of the
Aharonov-Bohm effect.

This work is a continuation of our previous article \cite{Santos.18},
where we proposed an strategy to optimize the correlation between the
extreme sites of a chain, while keeping the energy gap constant. The
optimal strategy found was, precisely, to increase certain hoppings,
presenting an intriguing similarity with the Su-Schrieffer-Heeger
(SSH) model of a dimerized fermionic chain
\cite{Su.79,Heeger.88,Sirker.14}, which is known to present a
non-trivial symmetry protected topological phase \cite{Asboth}. In
this case, the large correlations pattern is extended to several
sites, in more complex configurations. 

This article is organized as follows. Sec.~\ref{sec:model} presents
the model for independent fermions. In Sec.~\ref{sec:edgedimer} we
summarize the behavior of the edge-dimerized chains, taken from
\cite{Santos.18}. Superlattices built from edge-dimerized chains are
the focus of Sec.~\ref{sec:superlatt}. Afterwards, in
Sec.~\ref{sec:junction} we discuss junctions with a central ring,
which present in some cases a topological obstruction to the
detachment, which can be removed using a magnetic flux. Our
conclusions are summarized in the last Section.


\section{Model}
\label{sec:model}

Let us consider a chain of $L$ sites and $N_e$ independent spinless
Dirac fermions. A tight-binding Hamiltonian can be written in the
following way:
\beq
H=-\sum_i t_i \ c^\dagger_i c_{i+1} + h.c.
\label{eq:ham}
\eeq
where $c^\dagger_i$ is the creation operator at site $i$ and the $t_i$
are the local hopping amplitudes.  Unless otherwise stated, we will
consider $N_e=L/2$, i.e., half-filling.

The ground state (GS) of \eqref{eq:ham} can always be written as a
Slater determinant:

\beq
\ket|\Psi>=\prod_{k=1}^{N_e} b^\dagger_k \ket|0>,
\label{eq:slater}
\eeq
with $\ket|0>$ the Fock vacuum and $b^\dagger_k$ the creation
operators for the orbitals, given by a canonical transformation

\beq
b^\dagger_k = \sum_i U_{ki} \ c^\dagger_i,
\label{eq:canonical}
\eeq
where $U$ is the matrix that diagonalizes the hopping matrix, $T_{ij}
= t_i \ (\delta_{i,i+1} + \delta_{i,i-1})$, with eigenvalues
$\varepsilon_k$.  The energy gap of the system is given by the minimal
excitation energy, $\varepsilon_{(L/2)+1}-\varepsilon_{L/2}$.

The correlation matrix, defined as

\beq
C_{i,j}=\bra<\Psi| c^\dagger_i c_j \ket|\Psi>.
\label{eq:corr}
\eeq
allows us to compute the expectation value of any observable on any
state given by Eq. \eqref{eq:slater}, via Wick's theorem.  It can be
evaluated using the matrix $U$:

\beq
C_{i,j}=\sum_{k=1}^{N_e} \bar U_{ki} U_{kj}.
\label{eq:corr_slater}
\eeq

Entanglement properties of a generic block of the chain,
$B=\{i_1,\cdots,i_\ell\}$ (note that the sites $i_1,\cdots,i_\ell$ are
possibly disjointed), are always referred to the reduced density
matrix of $\ket|\Psi>$, defined as

\beq
\rho^B \equiv 
\Tr_B \ket|\Psi>\bra<\Psi|  ,
\label{eq:rho}
\eeq
being $\Tr_B$ the partial trace on the block. In the case of a Slater
determinant, this $\rho^B$ can be expressed as a tensor
product of $2\times 2$ density matrices of the form \cite{Peschel.03}

\beq
\rho^B = \bigotimes_{k=1}^\ell
\pmat{\nu^B_k & 0 \\ \\ 0 & 1-\nu^B_k}.
\eeq
where the $\nu^B_k \in [0,1]$ are the eigenvalues of the correlation
$\ell\times\ell$ sub-matrix corresponding to the block $C^B$ (i.e.,
those elements of $C_{i,j}$ with $i, j \in B$).

These eigenvalues can be interpreted as occupation numbers for
fermionic modes, associated to a set of {\em entanglement energies}
$\epsilon^B_k$ given by

\beq
\nu^B_k = 
\frac{1}{1+\exp \epsilon^B_k}.
\label{eq:es}
\eeq
These values $\{\epsilon^B_k\}$ constitute the so-called {\em
  entanglement spectrum} (ES) \cite{Li_Haldane.08}, which make up the
single-body energies of the {\em entanglement Hamiltonian} (EH),
$H^B$, defined as the Hamiltonian whose thermal state at $\beta=1$
corresponds to the actual reduced density matrix,

\beq
\rho^\mathcal{B}=\exp(-H^B).
\label{eq:eh}
\eeq
For free fermionic states, the EH is a quadratic Hamiltonian with a
new effective hopping matrix which can be obtained from the
correlation matrix \cite{Eisler.17}. The {\em entanglement gap} is
defined as the energy gap of $H^B$,
$\epsilon^B_{(L/2)+1} - \epsilon^B_{L/2}$.

On the other hand, the entanglement entropy of the block, defined as
the von Neumann entropy of $\rho^B$,

\beq
S^B = - \Tr \rho^B \log\rho^B,
\label{eq:Srho}
\eeq
can be computed using the following expression 
\cite{Peschel.03}

\beq
S^B = -\sum_{k=1}^\ell 
\Big( 
\nu_k^B \log\nu_k^B + 
(1- \nu_k^B) \log(1-\nu_k^B) 
\Big).
\label{eq:S}
\eeq

Our interest in the aforementioned entanglement measures stems from
the fact that they are usually able to characterize the different
phases of matter through e.g. corrections of (or violations to) the
area law for the entanglement entropy \cite{Vidal.03,Ramirez.14}.

Beyond the entropy, the ES and EH provide a complete characterization
of the dynamics of a block as a mixed state embedded in the ground
state of the total system.  The EH can be regarded as the effective
Hamiltonian describing the state of the block.  Thus, the EH of a
disjoint block can provide an estimate of the effective couplings
between different regions of the system.


\section{Edge-dimerized chain}
\label{sec:edgedimer}

\begin{figure*}
  \includegraphics[width=12cm]{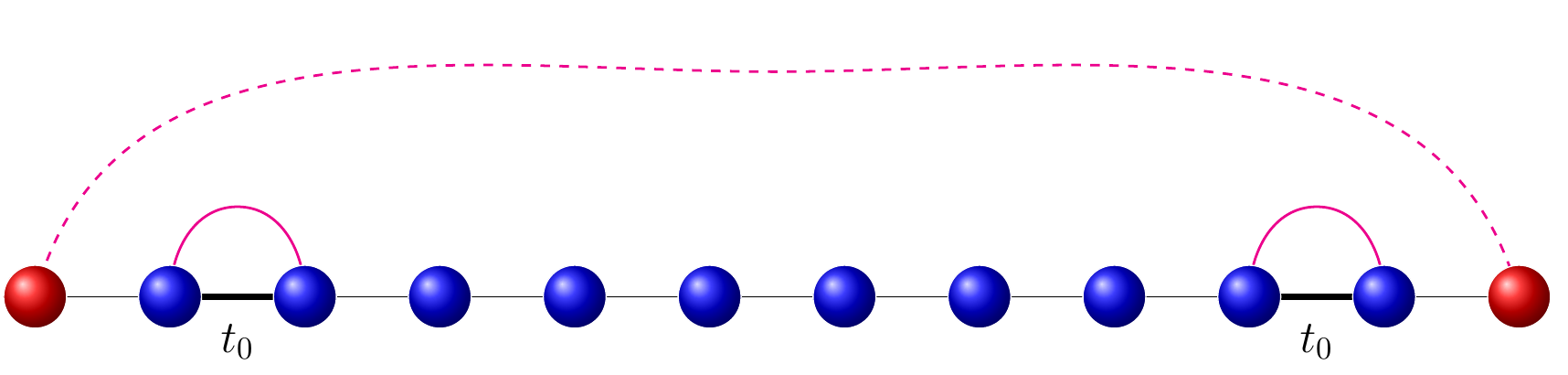}
  \caption{Illustration for the edge-dimerized chain. The second and
    penultimate hoppings are set to a different value than the rest,
    $t_0$. If $t_0\gg 1$, a valence bond is set on top of these links,
    thus forcing a long-distance valence bond between the first and
    last sites.}
  \label{fig:illust_edge}
\end{figure*}

Our first example of {\em entanglement detachment} is provided by the
{\em edge-dimerized} chain, described in \cite{Santos.18} as a
nearly optimal way to obtain large end-to-end correlations with a
robust energy gap. Consider an open fermionic chain of $L$ (even)
sites, subject to the Hamiltonian \eqref{eq:ham}, with the hoppings
$t_i$ given by

\beq
t_i=\begin{cases} 1 & \textrm{if } i\neq 2, L-1, \\
t_0 & \textrm{if } i=2 \textrm{ or } i=L-1, \\
\end{cases}
\label{eq:hoppings_edge}
\eeq
i.e.: only hoppings $t_2=t_{L-1}=t_0$ are distinguished, as
illustrated in Fig. \ref{fig:illust_edge}. If $t_0\gg 1$, a valence
bond will appear on top of links $2$ and $L-1$, i.e. between sites $2$
and $3$ and between sites $L-2$ and $L-1$, thus leaving sites $1$ and
$L$ no option but to establish a valence bond themselves. Otherwise,
if $t_0\ll 1$, the chain effectively decouples: the block containing
sites $\{1,2\}$ and the block containing sites $\{L-1,L\}$ become
isolated from the rest of the chain.
  
\begin{figure}
  \includegraphics[width=8.5cm]{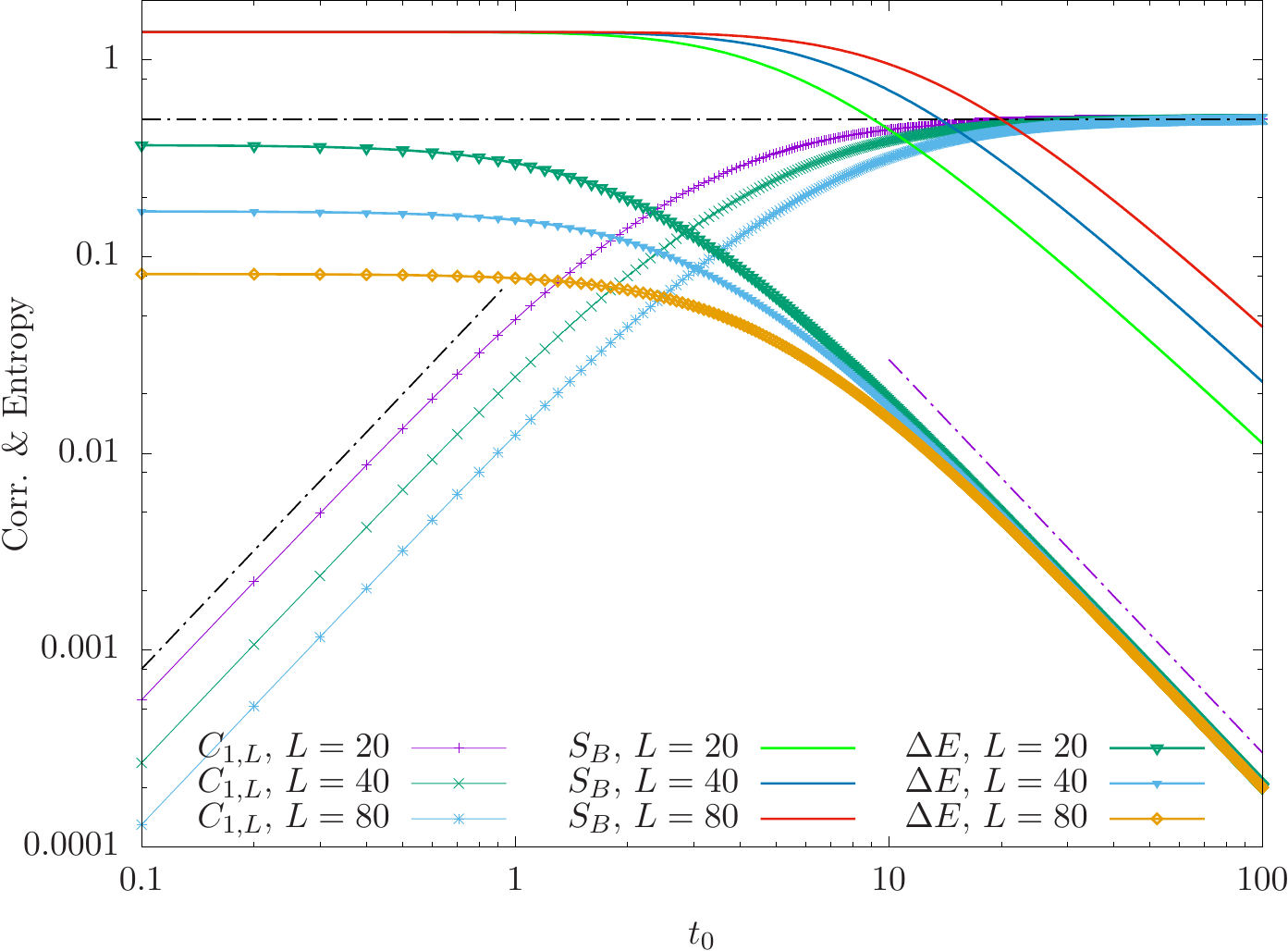}
  \caption{End-to-end correlation, $|C_{1,L}|$, energy gap, $\Delta
    E$, and entanglement entropy $S^B$ of the block $B=\{1,L\}$ as a
    function of $t_0$ in the Hamiltonian \eqref{eq:ham} with hoppings
    given by Eq. \eqref{eq:hoppings_edge}. The marked dashed lines
    correspond to scalings $t_0^2$ and $t_0^{-2}$. The horizontal line
    marks the limit correlation of $1/2$.}
  \label{fig:edge_S}
\end{figure}

Fig. \ref{fig:edge_S} shows the correlation between sites $1$ and $L$,
or end-to-end correlation, $|C_{1,L}|$, as a function of $t_0$ for
different values of $L$, within the GS of Hamiltonian
\eqref{eq:ham}. We can see that, for low $t_0$, the correlation grows
as $t_0^2$, as discussed theoretically in \cite{Santos.18}, while it
saturates to $1/2$ for $t_0\to\infty$. This large value of the
correlation implies that a valence bond has been created between these
two extreme sites. In other words, they have established a {\em Bell
  pair}, which is maximally entangled. Thus, through the monogamy of
entanglement, the block constituted by sites $B=\{1,L\}$ should detach
from the rest of the chain. In Fig. \ref{fig:edge_S} we can also see
the entropy of this block, $S^B$, as a function of $t_0$. Indeed, we
observe that for very low $t_0$, $S^B\approx 2\log 2$, which is the
maximal possible entanglement. As we increase $t_0$, the entropy
decreases steadily, tending to zero for large $t_0$. Thus, sites $1$
and $L$ become effectively detached, and the global wavefunction
becomes factorizable.

In Fig. \ref{fig:edge_S} we have also plotted the energy gap, $\Delta
E$ as a function of $t_0$ for different values of $L$. The energy gap
is a measure of the robustness of the ground state. We can observe
that the energy gap falls as $\Delta E\sim t_0^{-2}$, as argued in
\cite{Santos.18}.

This example serves as an illustration of the general phenomenon,
which will be discussed in more interesting situations in the
following sections: as a few couplings are {\em increased}, a certain
block may become {\em detached} from the rest of the system, i.e.: the
entanglement between the block and the rest of the system will fall to
zero. 


\section{Superlattice chain}
\label{sec:superlatt}

\begin{figure*}
  \includegraphics[width=16cm]{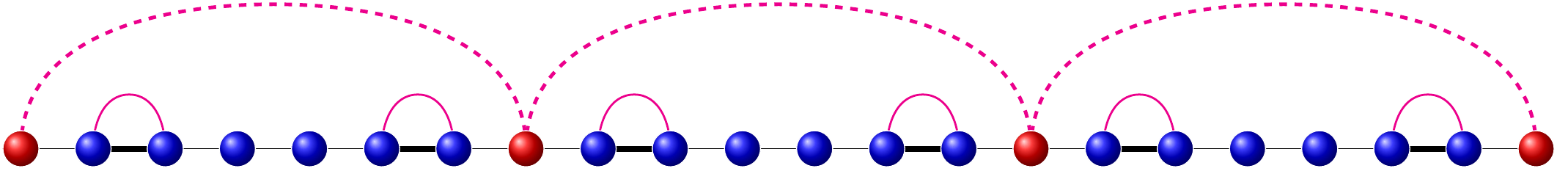}
  \caption{Illustration of the system used to detach a superlattice
    from the rest of the chain, inducing large multiparty correlations
    among the superlattice sites. The chain has periodic boundaries
    (the last site is the same as the first) and presents regular
    (thin black lines) and strong (thick black lines) links. The red
    sites, which make up the superlattice, are surrounded by sites
    which form strong bonds, thus effectively separating them from the
    rest of the chain. The strong links will tend to form a valence
    bond (red lines). Large distance correlations between the
    superlattice sites may appear (as indicated with dashed red
    lines). In this example, we have $m=3$ superlattice sites
    (remember that first and last sites are the same) separated by
    $\ell=7$ hoppings. The number of sites is $L = m \times \ell = 21$
    in this case.}
  \label{fig:superlatt}
\end{figure*}

\begin{figure}
  \includegraphics[width=8.5cm]{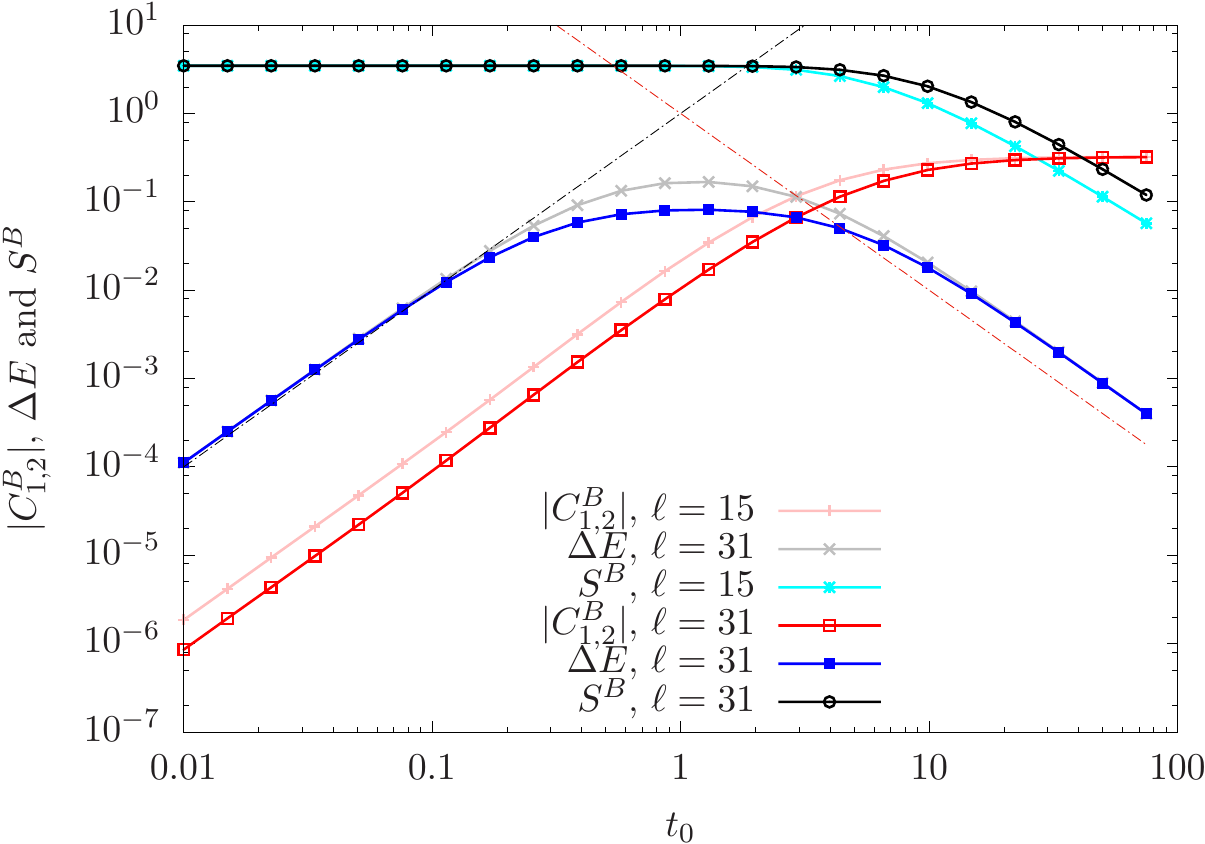}
  \caption{Superlattice system with $m=5$ and $\ell=15$ or $31$,
    i.e. with $L=75$ or $155$ sites. The enhanced hopping term is
    given by $t_0$. The correlation between neighboring superlattice
    sites (those in red in Fig.~\ref{fig:superlatt}) $|C^B_{1,2}|$
    increases like $t_0^2$ for low $t_0$, and saturate at a value
    $\sim 0.32$ for both system sizes. The energy gap, $\Delta E$, can
    be seen to grow as $t_0^2$ for low $t_0$ and decrease like
    $t_0^{-2}$ for large $t_0$, having a maximum at $t_0=1$. The block
    entropy of the superlattice sites, $S^B$ decreases from its
    maximum possible value for $t_0\to 0^+$, $5\log(2) \sim 3.46$, as
    $t_0^{-2}$ when $t_0>1$.}
  \label{fig:corr_sl}
\end{figure}

Let us introduce a more interesting situation, where entanglement
detachment takes place for a disconnected block making up a {\em
  superlattice} whose sites are periodically spaced along the
chain. As we will show, the superlattice can be effectively
disentangled from the rest of the system by enhancing a few selected
hoppings.

Let us consider a periodic chain of $L$ sites, with $L=\ell\times m$,
where $m$ will be the number of sites in the superlattice block and
$\ell$ will be the distance between them, as illustrated in
Fig. \ref{fig:superlatt} using $m=3$ and $\ell=7$. We will focus on
the case when both $m$ and $\ell$ are an odd numbers. Thus, $L$ is an
odd number and it is impossible to achieve a half-filling
occupation. We will then consider the system with $N_e=(L-1)/2$
fermions. The superlattice sites will be denoted by
$B=\{s_1,\cdots,s_m\}$, where $s_j=(j-1)m$.

The system will also be subject to Hamiltonian \eqref{eq:ham}, with
all hoppings set to $t_i=1$ except a few selected ones enhanced to
$t_i=t_0$. These selected hoppings are the second and penultimate of
each segment between superlattice sites, and they are denoted by thick
black lines in Fig. \ref{fig:superlatt}. When $t_0\gg 1$, these
hoppings are strong and a valence bond state tends to be placed upon
them, thus forcing the superlattice sites to live in relative
isolation and inducing them to establish large correlations among
themselves. Let us consider the submatrix $C^B$, denoted by

\beq
C^B_{i,j}=\<c^\dagger_{s_i}c_{s_j}\>,
\eeq
on the ground state. Notice that, due to the periodic boundary
conditions, $C^B_{i,j}$ only depends on $|i-j|$. Using
Eq. \eqref{eq:S} we can obtain the entanglement entropy between the
superblock and the rest of the system. As we will show, the
superlattice effectively detaches from the rest of the system for
large values of $t_0$.

We present in Fig.~\ref{fig:corr_sl} the correlation between
neighboring superlattice sites, $|C^B_{1,2}|$, as a function of $t_0$
for a system with $m=5$ and different values of $\ell$ (15 and 31).
We notice that if $t_0<1$ the correlation grows steadily as $t_0^2$,
and it saturates for both values of $\ell$ at the same value, close to
$|C^B_{1,2}|\sim 0.32$. Fig. \ref{fig:corr_sl} also provides
information about the energy gap, $\Delta E$, which can be seen to
decay both for large and small values of $t_0$, like $t_0^{-2}$ and
$t_0^2$ respectively, with a maximum at $t_0=1$.

The last magnitude shown in Fig. \ref{fig:corr_sl} is the superlattice
block entropy, $S^B$, associated to the partition between the
superlattice and the rest of the system. We can see that when $t_0
\lesssim 1$ the entropy $S^B$ takes its maximum possible value,
$5\log(2) \approx 3.46$ (because $n \log(2)$ corresponds to $n$ bond
cuts, see \cite{Ramirez.14}), showing that for such low values of
$t_0$ there is no connection between the superlattice sites. But for
$t_0>1$ the block entropy starts to decrease, also with a $t_0^{-2}$
behavior, marking the beginning of the slow detachment of the
superlattice block from the rest of the system.

Remarkably, the correlation, the gap, and the entropy of the block of
superlattice sites, are rather independent of the system size.
Moreover, we notice that the entanglement detachment, as measured by
the superlattice entropy, only takes place when the correlation
$|C^B_{1,2}|$ is nearly saturated.

\subsection{Entanglement Hamiltonian}

\begin{figure}
\includegraphics[width=8.5cm]{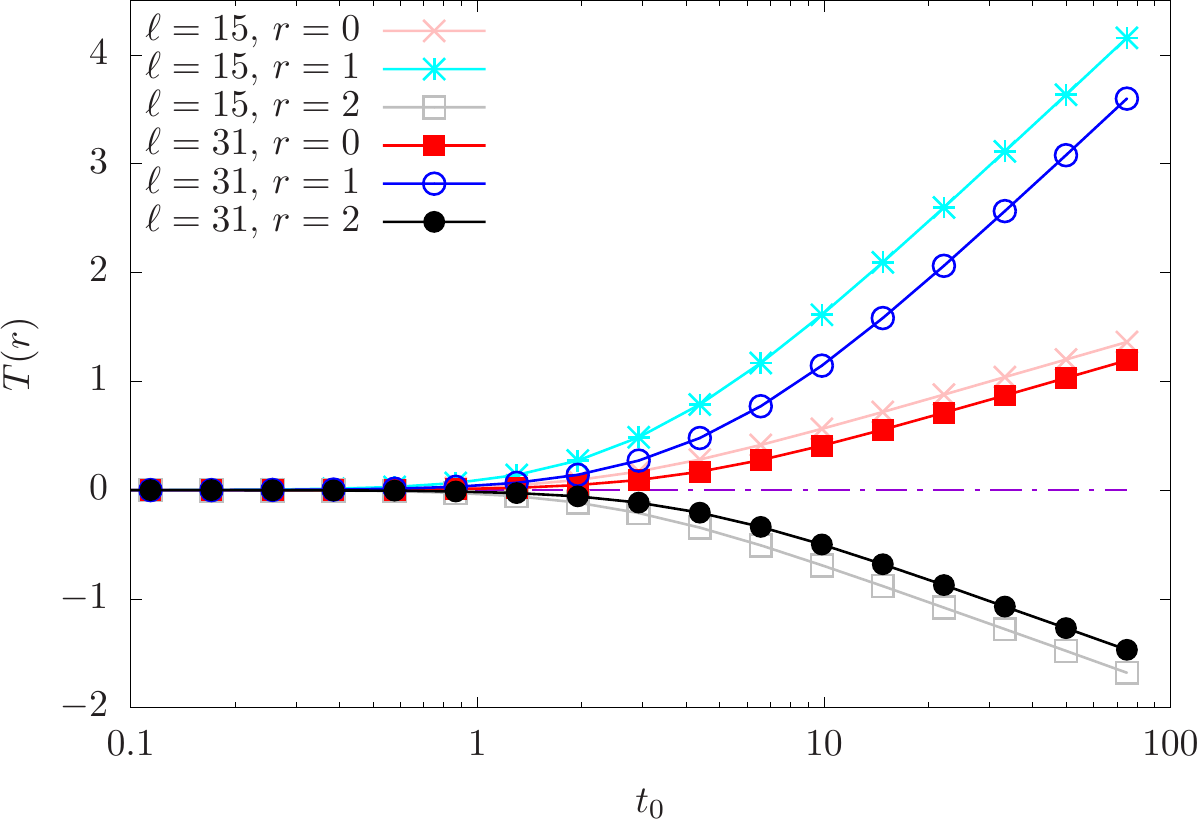}
\includegraphics[width=8.5cm]{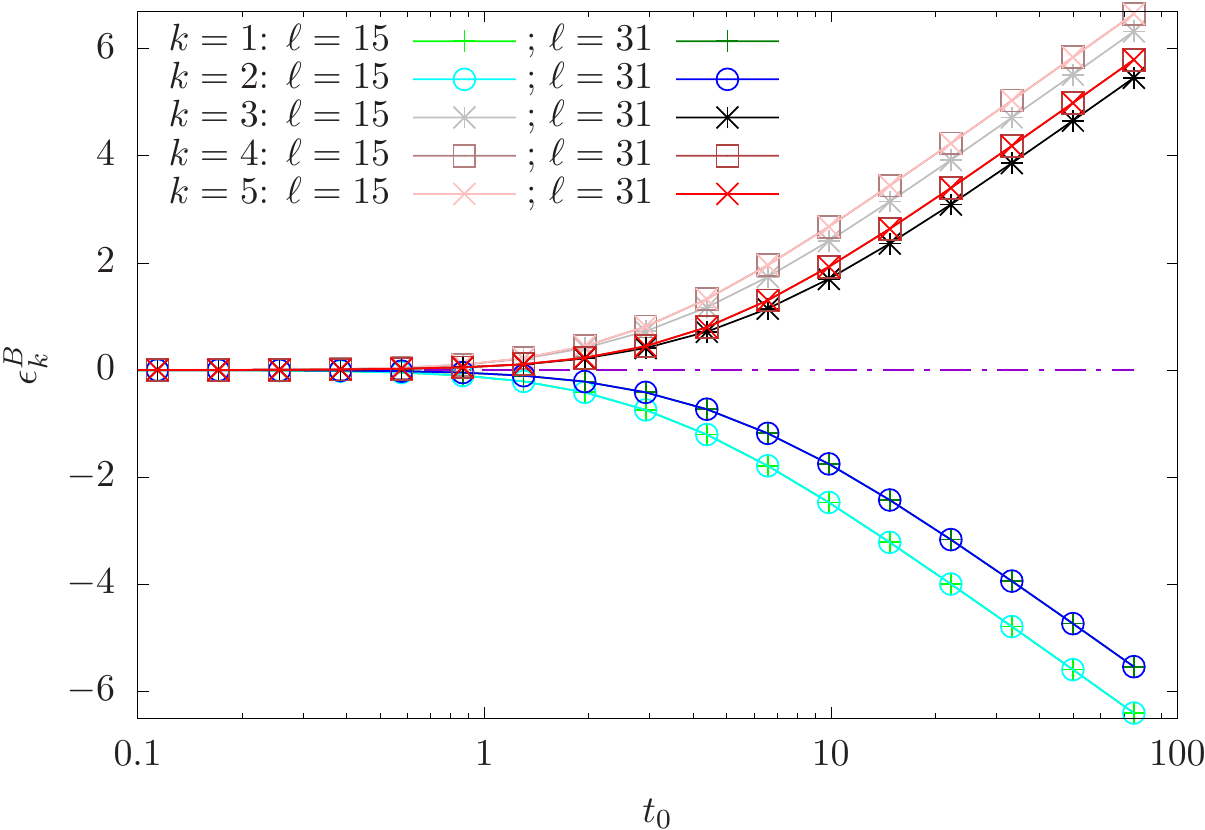}
\caption{Top: Entanglement Hamiltonian {\em effective hoppings},
  $T(r)$ with $r=|i-j|$ (see Eq. \ref{eq:EH_sl}) for the superlattice
  block with $m=5$ and $\ell=15$ and $31$, as described in
  Sec. \ref{sec:superlatt}. Notice that the effective hoppings grow
  logarithmically with $t_0$ for large values of $t_0$. Bottom:
  entanglement spectrum (ES) of the superlattice block,
  $\epsilon^B_k$, as a function of $t_0$ for the same two
  systems. Notice the exact degeneracy of the first two levels,
  $k=1,2$ and the last two, $k=4,5$.}
\label{fig:eh_sl}
\end{figure}

As the superlattice gets more and more detached, it becomes relevant
to ask what is the effective Hamiltonian describing the subsystem.  As
we know, the EH is an effective way to describe a certain subsystem
when we do not have access to the rest of the system, see
Eq. \eqref{eq:eh}. For a system of independent fermions, the EH is a
quadratic Hamiltonian in the fermionic fields
\cite{Peschel.03,Eisler.17,Tonni.18} and, due to the periodic
structure of our system, its elements must be translation invariant.
Thus, we can write

\beq
H^B = 
-\sum_{i,j}  T(|i-j|) \ c^\dagger_{s_i} c_{s_j} 
+ \mathrm{h.c.},
\label{eq:EH_sl}
\eeq
where $T(|i-j|)$ denote the {\em effective hopping} terms between the
superlattice sites (not only nearest neighbors), which would contain
the whole information of the superlattice subsystem. The top panel of
Fig. \ref{fig:eh_sl} shows these matrix elements, $T(|i-j|)$, as a
function of $t_0$ for the same superlattice system as before. Notice
that $T(0)$ plays the role of a chemical potential, while $T(1)$ is
the usual nearest-neighbor hopping term, while $T(r)$ for $r\geq 2$
become non-local hopping terms, which is non-negligible. All the terms
present a logarithmic growth for large $t_0$.

The entanglement spectrum (ES) also provides more complete information
about the entanglement structure \cite{Li_Haldane.08}. Indeed, it has
been widely used in order to detect non-trivial topological phases
\cite{Li_Haldane.08,Pollman.10,Ye.16}. The single-body ES of the
superlattice block is represented in the bottom panel of
Fig. \ref{fig:eh_sl}. The $\epsilon^B_k$, which constitute the
single-body spectrum of Hamiltonian \eqref{eq:EH_sl}, grow
logarithmically with $t_0$ for large $t_0$. The two lowest energy
levels and the two highest ones are exactly degenerate.

Notice that both the EH and the ES describe the reduced density matrix
of the superlattice block {\em as if} it corresponded to an
effective free-fermionic system at temperature
$\beta=1/(k_BT)=1$. But, alternatively, we may extract an effective
temperature as a prefactor for large values of $t_0$,

\beq
\rho^B(t_0)=\exp(-H^B(t_0))\; \sim\; \exp\(-\log(t_0)H^B_\eff\),
\label{eq:effH}
\eeq
where $H^B_\eff$ is expected to be independent of $t_0$, and
corresponds to a {\em limit Hamiltonian}. This scaling relation
considers the entanglement Hamiltonian as a fixed operator, and the
dependence on $t_0$ is attached to an {\em effective temperature},

\beq
k_B T_\eff = {1\over \log t_0},
\label{eq:teff}
\eeq
Thus, for very large $t_0$, $T_\eff\to 0$ and the superlattice becomes
disentangled from the rest of the system.


\begin{figure}
  \includegraphics[width=7cm]{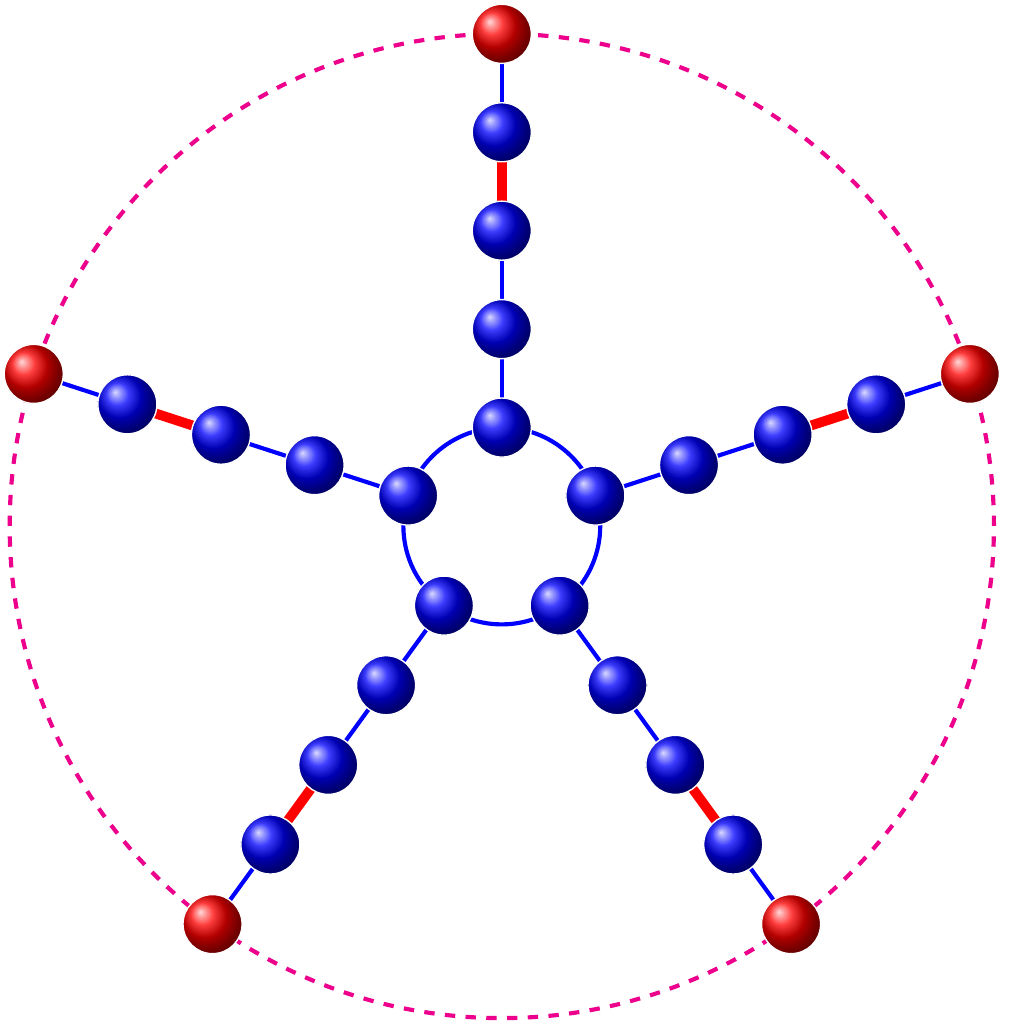}
  \caption{Illustration of a ring-junction with $n=5$ legs and
    $\ell=5$ sites per leg. The second link from the edge of all legs
    (in red) is enhanced, with a value $t_0>1$; all other links have a
    unit coupling. The edge sites (in red), even they are not
    physically connected, will establish large correlations among
    themselves (as indicated with red dashed lines).}
  \label{fig:junction_illust}
\end{figure}

\begin{figure}
  \includegraphics[width=8.5cm]{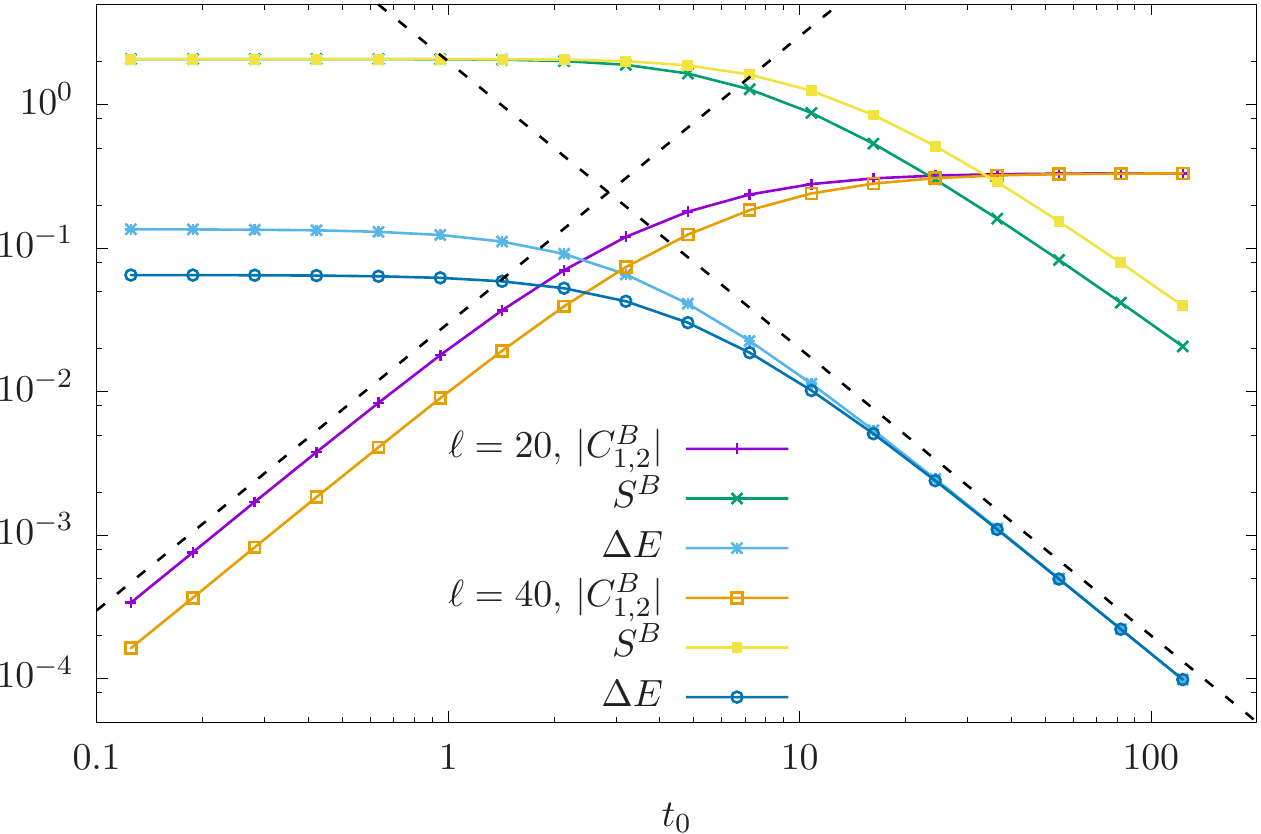}
\vspace{0.4cm}
  \includegraphics[width=8.5cm]{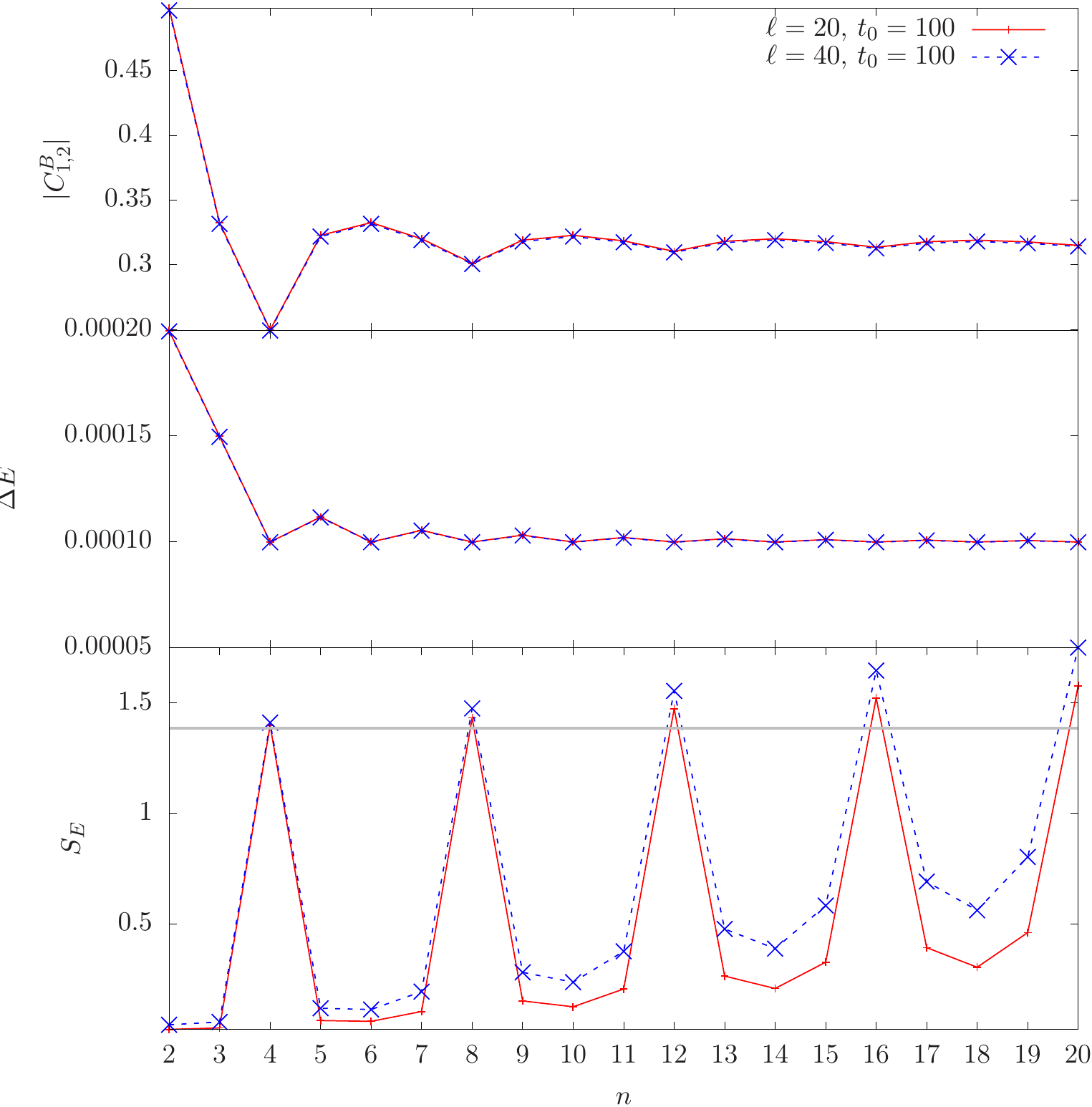}
\vspace{-0.7cm}
\caption{Top: For an edge-dimerized ring-junction with $n=3$,
  $\ell=20$ and $40$, we show the correlation between edge sites
  corresponding to neighboring legs, $|C^B_{1,2}|$, the entropy of the
  edge block, $S^B$, and the energy gap of the GS, all of them as a
  function of $t_0$. Notice the straight lines, which correspond to
  scalings $t_0^{-2}$ and $t_0^2$. Bottom: same magnitudes for
  edge-dimerized ring-junctions as a function of the number of legs,
  $n$, for a fixed value of $t_0=100$. Notice the periodic pattern in
  the entropy, $S^B$, where the horizontal line marks the value
  $2\log(2)$.}
\label{fig:junction}
\end{figure}

\section{Edge-dimerized Ring-Junctions}
\label{sec:junction}

Let us consider the ring-junction illustrated in
Fig. \ref{fig:junction_illust}. The fermionic sites are arranged in
$n$ chains of $\ell$ sites joined through one extreme, making up a
central ring. All hoppings will take unit value, except the second
link starting from the free edge (marked in red), which will take a
value $t_0$. Now let us consider the block formed by the free extremes
of each leg (marked in red), $B=\{s_1,\cdots,s_n\}$. We will show that
this {\em edge block} detaches from the rest of the system when $t_0$
becomes very large, while it establishes large internal correlations
despite the lack of physical hoppings among them.

Please note that the sites making up the edge block $B$ (marked in red
in Fig. \ref{fig:junction_illust}) are {\em physically
  disconnected}. For a particle to jump from one to any other, it must
hop along one leg in order to reach the center, and then along another
leg, in order to reach the new extreme, making a total of $2\ell-1$
hoppings. Nonetheless, as we shall show, {\em effective} hopping terms
between them will appear in the entanglement Hamiltonian.

Let $C^B$ be the correlation sub-matrix corresponding to the edge
block, such that $C^B_{1,2}$ stands for the correlation between edge
sites of neighboring legs. The top panel of Fig. \ref{fig:junction}
shows the dependence of this correlation $|C^B_{1,2}|$ on the
designated hopping, $t_0$, for a ring-junction with $n=3$ legs and two
leg sizes: $\ell=20$ and $\ell=40$. We observe, as in the other cases,
the $t_0^2$ dependence for low $t_0$, with a saturation at a finite
value a bit less than $1/2$. The top panel of Fig. \ref{fig:junction}
also shows the energy gap, $\Delta E$ and the block entropy $S^B$ as a
function of $t_0$ for the same systems. We notice these two magnitudes
decrease steadily for large $t_0$. Thus, we make our main claim: the
{\em edge of the ring-junction detaches} for large values of $t_0$.

A very relevant effect shows up when we consider different values for
the number of legs, $n$, as shown in the bottom panel of
Fig. \ref{fig:junction} using always $t_0=100$, which is close enough
to the strong coupling limit, when $t_0\to\infty$. The top plot shows
the behavior of the correlation $|C^B_{1,2}|$, where we observe
oscillations with $n$ which attenuate for large $n$, around a value
$|C^B_{1,2}|\sim 0.314$, i.e., quite high. The energy gap, $\Delta E$,
is shown in the second plot, also as a function of $n$. In this case
we also observe some oscillations which attenuate for large $n$ around
a limit value. The most intriguing situation shows up when we study
the entanglement entropy of the edge block, $S^B$, as shown in the
third plot. There we can see that the oscillation becomes much more
marked, and does not attenuate. Besides a certain secular
upwards trend, all values of $n$ which are multiples of $4$ present a finite
entanglement entropy $S^B$ which is close to $2\log(2)$ (marked with
the horizontal line). Thus, we observe that, if the number of legs
$n=4k$, the edge block {\em does not detach}.

\begin{figure}
\includegraphics[width=8.5cm]{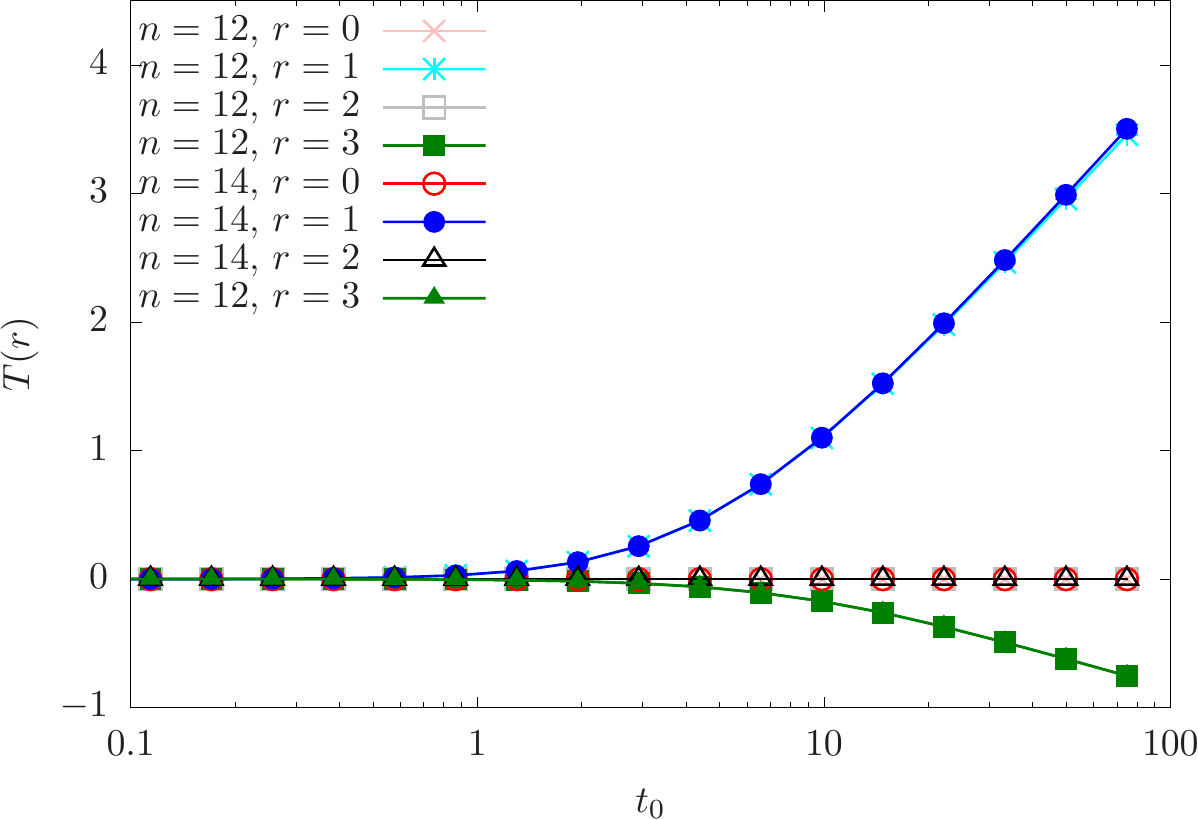}
\includegraphics[width=8.5cm]{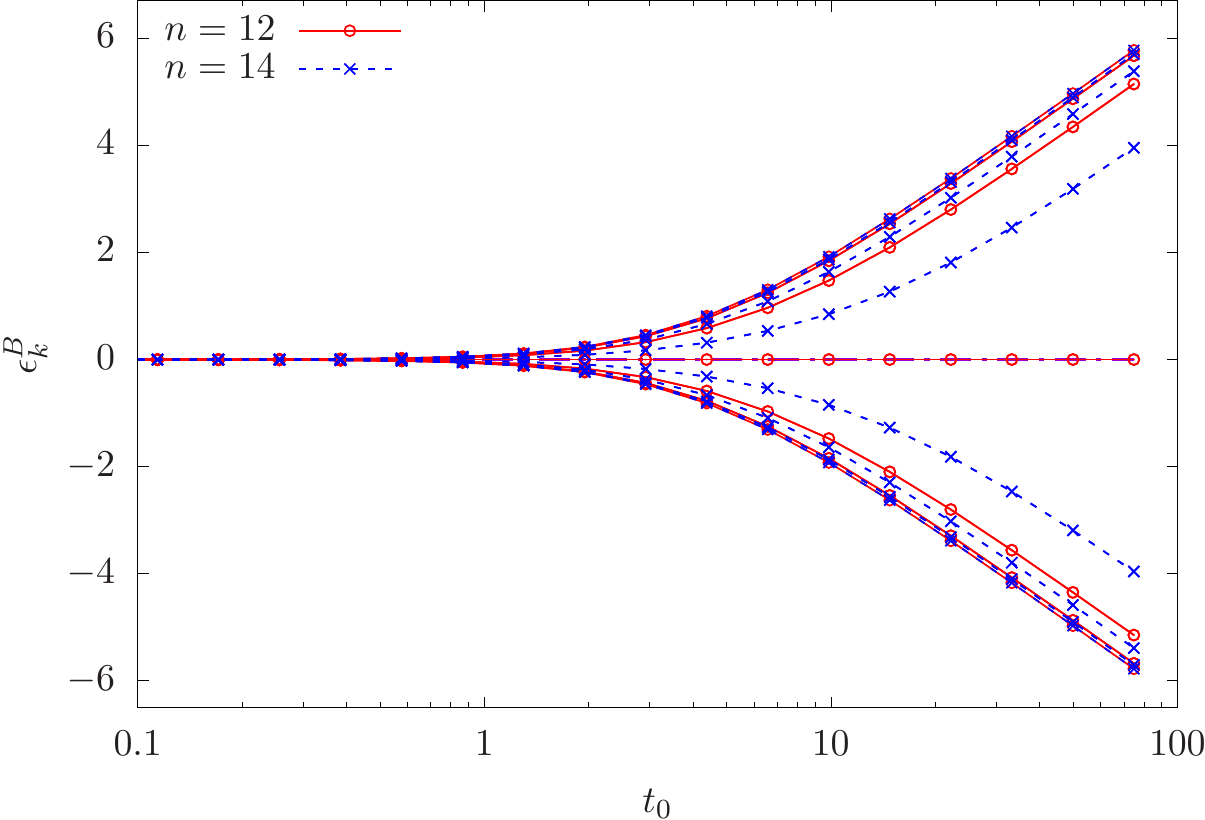}
\caption{Top: Entanglement Hamiltonian coefficients, $T(r)$, for
  edge-dimerized ring-junctions with $\ell=40$ sites per leg, $n=12$
  and $n=14$ legs, as a function of $t_0$ for the first few values of
  $r$. Notice that, for large $t_0$, they grow like
  $\log(t_0)$. Moreover, the coefficients present very small
  differences from $n=12$ to $n=14$. Bottom: Entanglement spectrum of
  the same systems. Notice the same scaling behavior, {\em with one
    exception}: the ES of the case $n=12$ presents two zero modes,
  $\epsilon^B_{n/2}=\epsilon^B_{n/2+1}=0$.}
\label{fig:junction_EH}
\end{figure}

In order to explain this unexpected periodic behavior, we will
consider the entanglement spectrum (ES) and entanglement Hamiltonian
(EH). We will assume that the EH presents the same form provided for
the superlattice, Eq. \eqref{eq:EH_sl}, since it must be a
free-fermion Hamiltonian with translation invariance. In the top panel
of Fig. \ref{fig:junction_EH} we show $T(r)$ for the first values of
$r$ using two systems with $\ell=40$ and two different numbers of
legs, $n=12$ and $n=14$, as a function of $t_0$. The most salient
feature is that $T(0)=T(2)=0$. All even terms are zero. Moreover, we
can observe that, for large $t_0$, they grow like $\log(t_0)$, as it
was the case for the superlattice, see Sec. \ref{sec:superlatt}.

The bottom panel of Fig. \ref{fig:junction_EH} presents the
entanglement energies, $\epsilon^B_k$ for the same systems. The
scaling behavior is very similar, they grow as $\log(t_0)$ for large
$t_0$. Yet, there is a crucial difference: the case $n=12$ presents an
exact {\em zero mode}, i.e. $\epsilon^B_{n/2}=\epsilon^B_{n/2+1}=0$
exactly. The same is true for all other $n=4k$.

Thus, we can still make similar claims to the superlattice case: there
is a scaling entanglement Hamiltonian, $H^B_\eff$, such that, for
large enough $\log(t_0)$,

\beq
H^B(t_0) \sim \log(t_0)\,H^B_\eff,
\eeq
But, interestingly, when $n=4k$, the GS of $H^B_\eff$ {\em presents
  exact degeneracy} of the ground state. This is a hallmark of a
non-trivial {\em topological phase} \cite{Pollman.10,Ye.16}, leading
to a minimal entanglement entropy, which is given by

\beq
S_\min = d \log(2),
\label{eq:degS}
\eeq
where $d$ is given, in our case, by the number of zero modes. Thus,
$S_\min = 2\log(2)$, in correspondence with the observed behavior of
the entanglement entropy for large $t_0$ in
Fig. \ref{fig:junction}. In other terms: for $n=4k$ the edge block
{\em does not detach} because it presents a minimal entropy of
topological origin.

\bigskip

This strong periodicity effect on the number of legs presents an
intriguing relation to H\"uckel's rule for {\em aromaticity}, which
establishes that for planar rings of carbon atoms, only those with
$4k+2$ atoms can be stable, due to the fact that $\pi$ orbital
electrons can be delocalized only if the plane ring has $4k+2$ $\pi$
electrons.  In our edge-dimerized junctions the edges will be isolated
or not from the rest of the system depending on the aromaticity of the
external effective ring of edge sites isolated for $4k+2$ legs
(aromaticity) and for odd number of legs (no aromaticity), but will
not be isolated for $4k$ legs (antiaromaticity). This aromaticity
effect is ubiquitous in ringed systems.
\cite{Zilberg.99,Wiberg.01,Pierrefixe.08,Breslow.14}


\subsection{Magnetic Flux through Junction Rings}
\label{sec:dhva}

The described behavior of the ring junction can change dramatically in
the presence of an external magnetic field. Let $\Phi$ be the magnetic
flux traversing the central ring, in units of the elementary quantum
flux $\Phi_0$. The magnetic flux causes the hopping amplitudes to
acquire a complex phase, such that the product of all phases computed
clockwise will be $\exp(2\pi i \varphi)$. Thus, we will simulate the
effect of the magnetic field by including a phase $\exp(2\pi i
\varphi/n)$ in each of the couplings along the ring. Physical
properties of the GS will be therefore periodic in the external flux
$\Phi$, in analogy with the de Haas-van Alphen effect \cite{Haas.30}.
We will consider the effect of the magnetic flux on the edge sites,
therefore an analogy with the Aharonov-Bohm effect can also be
established \cite{Orellana.03}.

\begin{figure}
  \includegraphics[width=8.5cm]{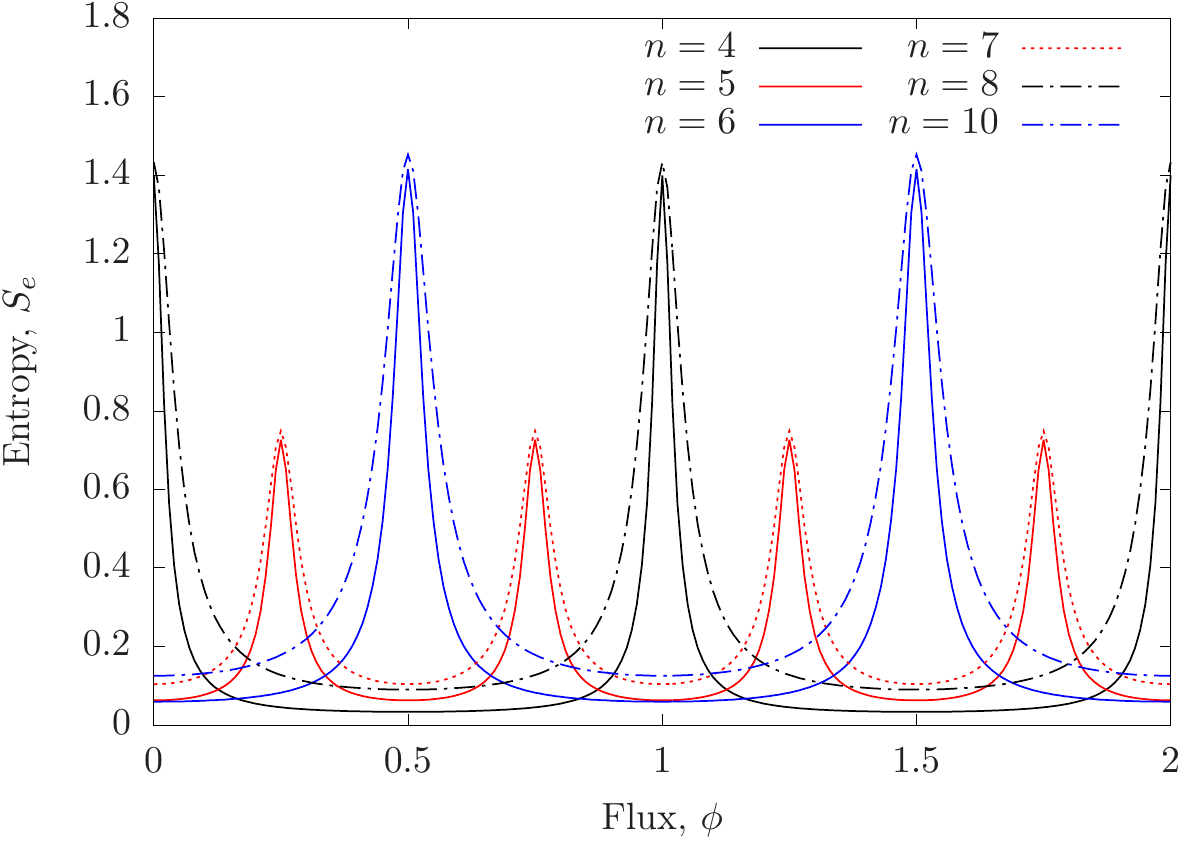}
  \caption{Analogue of de Haas-van Alphen effect (dHvA) in junction
    rings. Entropy of the edge ring, $S^B$, as a function of the flux
    $\Phi$ traversing the central ring. We have used $\ell=20$ sites
    per leg, and $n=4$, 5, 6, 7, 8 and 10 legs. The maxima of the
    peaks correspond to $\log(2)$ and $2\log(2)$.}
  \label{fig:dhva}
\end{figure}

Fig. \ref{fig:dhva} shows the entropy of the edge block, $S^B$, as a
function of the magnetic flux crossing the central ring, for $\ell=20$
and different numbers of legs, $n$. The periodicity in $\Phi$ is
perfect. Moreover, the entropy presents strong peaks at different
values of the magnetic flux, which depend only on the value of $n
\inmod 4$. For $n=4k$, the {\em antiaromatic} junctions, we see that
the entropy peaks appear at integer values of the flux $\Phi$. For
$n=4k+2$, corresponding to the {\em aromatic} junctions, the entropy
peaks appear at half-integer values of the flux. In these two cases,
the maximal entropy is close to $2\log(2)$. For odd $n$, the peaks
appear at $\Phi=k+1/4$ and $k+3/4$, for any integer $k$, but the
maximum value is only $\log(2)$.

The physical interpretation is as follows. The GS of the EH presents
zero modes for special values of the external flux, which depend on
the value of $n \inmod 4$.  For odd values of $n$, one zero mode
appears at $\Phi=1/4$ and the other at $\Phi=3/4$. Thus, the
degeneracy of the GS of the EH is double and $S^B$ peaks at
$\log(2)$. For even $n$, the zero modes coincide, but they appear at
different values of the magnetic flux, thus increasing the minimal
entanglement entropy to $2\log(2)$, according to Eq. \eqref{eq:degS}.

In other words, for large $t_0$ the ring-junction system develops an
edge Hamiltonian which is degenerate at specific values of the
magnetic flux. Although the reduced density matrix can always be
considered to be close to zero temperature, the degenerate entanglement
spectra force a minimal entanglement entropy of topological origin
$S_\min=d\log(2)$, where $d$ is the number of zero modes
encountered. Thus, we see that the entropy of the edge block of
ring-junctions develops a high sensitivity to the magnetic flux $\Phi$
that traverses their rings.


\section{Conclusions and further work}
\label{sec:conclusions}

In this article we have explored simple quantum systems for which an
increase in a few selected couplings can lead to the detachment of
some block, as signaled by a vanishing entanglement entropy between
the block and the rest of the system. The sites corresponding to the
block need not be neighbors, and they will develop large
correlations among them. We have described a few systems based on
free-fermions on different quasi-1D geometries which present this {\em
  entanglement detachment}.

Our first example was provided by an open free fermionic chain where
the second and penultimate hoppings are increased, thus detaching the
block formed by the first and last sites. In the second example we
detached a {\em superlattice} out of a 1D closed fermionic chain by
enhancing selected hoppings. In the third and most elaborate example,
we displayed the fermionic sites along a {\em ring-junction}, a star
graph with a central ring, enhancing the penultimate hopping of each
leg.

We have characterized the reduced density matrices of the nearly
detached blocks via the entanglement entropy, entanglement spectrum
(ES) and Hamiltonian (EH). This way we could conclude that, both in
the superlattice and the ring-junction, the EH present a simple
scaling form for large, yet finite, enhanced hoppings. The case of the
ring-junction presents an intriguing periodicity in the number of
legs $n$: {\em anti-aromatic} junctions with $n=4k$ do not display
this entanglement detachment. Instead, they have a minimal
entanglement entropy equal to $2\log(2)$. We have shown that this
minimal entanglement entropy can be understood as an effect of the
presence of two zero modes in the ES, which signal a non-trivial
topological phase.

The topological effect becomes more interesting when we introduce an
external magnetic flux through the central ring. We can observe that
the entropy presents peaks of amplitude $S=d\log(2)$ with $d=1$ or
2. The position and height of the peaks depend on the number of legs
mod 4, and can be understood in terms of the zero modes of the ES and
the associated topological entropy.

The systems studied can be engineered in the laboratory. These types
of hopping patterns can appear naturally in quantum wires
\cite{Ahn.03,Ahn.05} or organic molecules \cite{Gruner.88}, or can be
engineered using optical lattices employing the so-called cold-atom
toolbox \cite{Lewenstein,toolbox}. This technology is also useful,
e.g., to simulate the effects of curved space-time on quantum matter
\cite{Celi.10,Laguna_Celi.17}.

Our work, thus, presents a proof-of-principle study of this
entanglement detachment, but many questions are still open: the
theoretical explanation of the scaling form of the entanglement
Hamiltonian, the possible extension to higher dimensions, the
applicability for quantum information processing and the study of
dynamical effects.


\begin{acknowledgments}

We would like to acknowledge very useful discussions with
S.N. Santalla and G. Sierra. J.R.-L. acknowledges funding from the
Spanish Government through Grant No. FIS2015-69167-C2-1-P.

\end{acknowledgments}


\end{document}